\documentclass[conference]{IEEEtran}
\IEEEoverridecommandlockouts

\usepackage{cite}
\usepackage{dsfont}

\usepackage[cmex10]{amsmath}

\usepackage{multirow}
\usepackage{booktabs,colortbl}

\usepackage[cmex10]{amsmath}
\usepackage{siunitx}
\usepackage{cite}
\usepackage{psfrag}
\usepackage[utf8]{inputenc}
\usepackage[T1]{fontenc}
\usepackage{amsmath,amsfonts,amsbsy,amssymb}
\usepackage{mathabx}
\usepackage{mathrsfs}
\usepackage[nolist]{acronym}
\usepackage{tabularx}
\usepackage{amssymb}
\usepackage{amsmath}
\usepackage{graphicx}
\usepackage{epstopdf}
\usepackage{cite}
\usepackage{multirow}
\usepackage{wasysym}
\usepackage{multirow}
\usepackage{float}
\usepackage{color}
\usepackage{algorithmic}
\usepackage{textcomp}
\usepackage[font=small,skip=0pt]{caption}

\newtheorem{proposition}{Proposition}
\newtheorem{lemma}{Lemma}

\newtheorem{remark}{Remark}
\def\BibTeX{{\rm B\kern-.05em{\sc i\kern-.025em b}\kern-.08em
		T\kern-.1667em\lower.7ex\hbox{E}\kern-.125emX}}

\begin{document}

\title{Increasing the Throughput of an Unlicensed Wireless Network through Retransmissions}

%
%
%
%
%
%
%
%

\author{\IEEEauthorblockN{Iran Ramezanipour\IEEEauthorrefmark{1}, Pedro. H. J. Nardelli\IEEEauthorrefmark{2}, Hirley Alves\IEEEauthorrefmark{1} and Ari Pouttu\IEEEauthorrefmark{1}}
	\IEEEauthorblockA{\IEEEauthorrefmark{1}Centre for Wireless Communications (CWC), University of Oulu, Finland\\
	}
	\IEEEauthorblockA{
		\IEEEauthorrefmark{2}Lappeenranta University of Technology, Lappeenranta, Finland\\
	}
	Firstname.lastname@oulu.fi, Firstname.lastname@lut.fi 
}


%

\maketitle

\begin{abstract}
This paper analyzes the throughput of an unlicensed wireless network where messages decoded in outage may be retransmitted.
We assume that some wireless devices such as sensors are the unlicensed users, which communicate in the licensed uplink channel.
In this case, the licensed users that interfere with the unlicensed transmissions devices are mobile devices whose spatial distribution are assumed to follow a Poisson point process with respect to a reference unlicensed link.
We investigate how the number of allowed retransmissions and the spectrum efficiency jointly affect the throughput in [bits/s/Hz] of a reference unlicensed link for different licensed network densities, constrained by a given required error rate.
The optimal throughput is derived for this case as a function of the network density.
We also prove that the optimal constrained throughput can always reach the unconstrained optimal value.
Our numerical results corroborate those of the analytical findings, also illustrating how the number of allowed retransmissions that leads to the optimal throughput changes with the error rate requirements.
\end{abstract}


\section{Introduction}

During the past decade, with the advancements in technology, more and more devices are being connected to the Internet each day \cite{iotserv}. 
The number of IoT enabled devices is estimated to reach hundreds of billions globally by the end of 2020. 
IoT is modernizing the way applications, services and humans are communicating with each other, covering broad areas such as transportation, public safety, healthcare, industries and housing. 
%

Before being incorporated under the IoT umbrella, wireless sensor networks (WSNs) have been for some time implemented to monitor their corresponding environments.
Now with widespread connections to internet, the sensor nodes as a part of IoT are able to join the internet when needed. 
However, the integration of WSNs in the IoT will open up new challenges as shown in \cite{iotserv}. We are going to witness a massive network of devices needing to join the internet and wireless networks during the upcoming years which is going to create several challenges for the current networks in terms of different network properties such as spectrum and bandwidth. 
Among the most urgent problems, the availability of the frequency spectrum and its current allocation policies impose challenges in both technical and political domains (refer to \cite{ak,pdnd} and references therein).

The general idea of cognitive radio tries to use the available spectrum more efficiently via adaptive solutions including, for instance, dynamic spectrum access. 
Within such a broad scope, we choose to focus on the concept of unlicensed spectrum access where unlicensed users share the same frequency band with the licensed users.
As discussed in \cite{pedro2,tome}, this choice makes sense for low-power IoT-based systems that employ the licensed uplink channel.
In a nutshell, the authors exploit the fixed position of the unlicensed users to justify the use of highly directional antennas in the unlicensed links with limited transmit power.
In this case, the interference from the unlicensed network to the licensed may be neglected.
On the other hand, the unlicensed users of the uplink channel experience the interference from the mobile users of the licensed network.

In \cite{pedro2,tome}, however, the study was centered on a specific smart grid application, limiting its possible usages.
Knowing this, we generalize here those ideas by extending the scope of the proposed scenario.
But, more importantly, we follow \cite{mar,pedro+marios} and include the possibility of retransmitting messages decoded in outage.
In those papers, the authors showed that allowing for a limited number of retransmissions may increase the transmission capacity \cite{mar} and the spatial throughput \cite{pedro+marios} of ad hoc networks.

Here, we adapt the previous system model to analyze the link throughput with focus on optimizing such a metric by jointly setting the spectral efficiency and the allowed retransmissions.
Our main contributions are:
\begin{itemize}
	\item We derive the number of allowed retransmissions and the spectrum efficiency that leads to the maximum link throughput for a reference unlicensed link;
	\item We show that the number of allowed retransmissions that maximizes the link throughput changes with the density of interferers and the error rate requirement;
	\item We prove that the unconstrained maximum throughput is always achievable regardless of the outage constraint.
\end{itemize}


\section{System model}
\label{sec:model}
%
We consider an extension to the scenario initially introduced in \cite{pedro2,tome} to deploy the communication network in which the sensors transmit their data to their corresponding aggregator/controller.
Specifically, we assume any application that the following assumptions hold.

\begin{itemize}
\item \textbf{Assumption 1}: Unlicensed spectrum access scenario where licensed and unlicensed/specially licensed users share the frequency bands allocated to the uplink channel.
\item \textbf{Assumption 2}: Licensed link is established between static cellular base-stations and mobile users. Unlicensed users are sensors that send data to a controller entity through the uplink channel, and their positions are fixed.
\item \textbf{Assumption 3}: Sensors transmit with limited power. The maximum power allowed for the unlicensed users can be seen as an imposition from the licensed network or related to the sensors' own capabilities.
\item \textbf{Assumption 4}: Packet collisions between sensors associated with the same aggregator/controller are neglected since the transmitted messages are assumed to be small and multiple access solutions are effective for the size of the unlicensed network.
\end{itemize}

These assumptions allow for some simplifications that are needed to carry out our analysis, as discussed next.
Assumption 2 states that the positions are fixed so the use of directional antennas in the unlicensed links is feasible as far as orientation errors can be eliminated~\cite{joint}.
Assumption 3 indicates the maximum range that the sensors' signal can reach.
Then, the radiation pattern created by the unlicensed transmission can be considered a line segment starting at the sensors to a maximum related to the maximum transmit power.

Based on Assumptions 1, 2 and 4, the co-channel interference happens: (i) from mobile users to aggregators/controllers, (ii) from sensors to cellular base-stations, (iii) from sensors to aggregators/controller that are not associated with each other.
From what was previously explained, it is possible to eliminate the interference in cases (ii) and (iii) by considering specific locations when implementing licensed/unlicensed networks. 
Even if the positions are randomly chosen, the probability that the base station or the controller are in the same line segment related to the sensor signal approaches zero.

The only relevant interference in this network is the case (i), which requires capturing the uncertainty of the active mobile users' positions  to characterize the effect of this interference on the system performance. 
To do so, the interfering nodes are modeled to follow a Poisson point process $\Phi$ distributed over an infinite two-dimensional plane with spatial density $\lambda$ (a more detailed explanation about this way of modeling can be found in \cite{haen}).


The wireless channel is modeled considering distance-dependent path-loss and fast fading.
Let $r_i$ be the distance between the reference receiver and the \textit{i}th interferer and $g_i$ to be the channel gain between them.
Then, the received power at the reference receiver is given by $W g_ir_i^{-\alpha}$, where $W$ is the transmit power and $\alpha> 2$ is the path-loss exponent. 
The signal to interference ratio SIR$_0$ at the reference receiver is:
\begin{equation} 
\label{eq_SIR}
\textrm{SIR}_0 = \dfrac{W_\mathrm{s} g_{0}  r_0^{-\alpha}}{W_\mathrm{p} \; \underset{i \in \Phi}{\displaystyle \sum}  g_{i}  r_i^{-\alpha}},
\end{equation}
where $W_\mathrm{p}$ and $W_\mathrm{s}$ are the transmit power employed by the licensed (interferers) and unlicensed (reference) transmitters.
Note that the noise is neglected here, but this assumption comes without significant qualitative differences, as discussed in  \cite{weber}.

\begin{figure}[!t]%
	
	\centering
	\includegraphics[width=	\columnwidth, height=4cm]{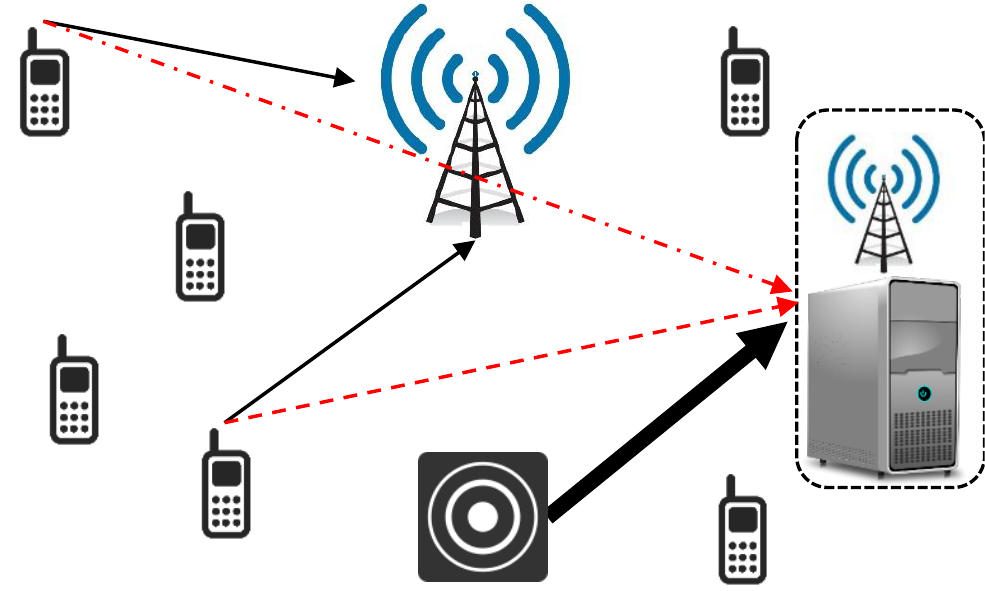}
	%
	\caption{An illustration of the proposed scenario, where licensed and unlicensed users share the up-link channel. The reference sensor (unlicensed transmitter) is depicted by the sensor, the controller (unlicensed receiver) by the CPU and its antenna, the handsets are the mobile licensed users (interferers to the controller) and the big antenna is the cellular base-station. As the sensors uses directional antennas with limited transmit power (bold arrow), its interference towards the base-station can be ignored. The thin black arrows represent the licensed users' desired signal, while the red ones represent their interference towards the controller.}
	\label{fig-scheme}
	\vspace{-3mm}
\end{figure}

The reference link employs point-to-point Gaussian codes and interference-as-noise decoding rules~\cite{ptp,pdnd} so a spectral efficiency of $\log_2(1+\beta)$ in bits/s/Hz can be attained only if the signal to interference ratio (SIR) is greater than a given threshold $\beta$ (i.e $\textrm{SIR}>\beta$).
An outage event occurs whenever $\textrm{SIR}\leq\beta$.
The probability that such an event occurs is denoted by $P_\textrm{out}$.
If recurrent outage events happen, the message can be retransmitted up to $m$ times \cite{mar}.
If the message is not successfully decoded after $1+m$ transmission attempts (one transmission plus $m$ retransmissions), it is dropped. 
The probability that the message is successfully decoded is then $P_\textrm{suc} = 1 - P_\textrm{out}^{1+m}$.

To compute  $P_\textrm{out}$, quasi-static channel gains (squared envelopes) $g$ are assumed. 
These gains are independent and identically distributed exponential random variables (Rayleigh fading) with mean $1$ \cite{joint}. 
The interfering nodes (i.e. licensed mobile users) are considered highly dynamic so their positions change at every transmission attempt by the reference link \cite{pedro+marios}. 
In this case, the signal-to-interference ratio at the reference link SIR$_0$ can be statistically evaluated by considering different realizations of Poisson point process $\Phi$. Every time a new packet is sent, there is a new realization of the network, hence, the transmissions are i.i.d.
The outage probability  $P_\textrm{out} = \textup{Pr}\left[\textup{SIR}_0 \leq \beta\right]$ for each transmission attempt is \cite{mar}:
\begin{equation}\label{eq:6}
P_\textrm{out}= 1 - e^{- k \lambda  \beta^{2/\alpha}},
\end{equation}
where $k=\pi r_0^2 \Gamma{\left(1 - \frac{2}{\alpha} \right)} \Gamma{\left(1 + \frac{2}{\alpha} \right)}$.

We can then compute the link throughput $T$ in the reference link including the possibility of $m$ retransmissions as \cite{mar}: 
\begin{equation}\label{eq:5}
T=\frac{\log(1+\beta)}{1+\bar{m}}\left(1-P_\textrm{out}^{1+m}\right),
\end{equation}
where $1+\bar{m}$ is the average number of transmissions needed to successfully transmit a message; note that $\left(1-P_\textrm{out}^{1+m}\right)$ is the probability that a message is successfully transmitted.

\section{Throughput optimization}
\label{sec:opt}

\subsection{Constrained optimization}
In this subsection, we assume a throughput optimization problem where the application imposes a maximum acceptable error rate as a quality requirement (i.e. how frequently a message is dropped after the allowed retransmissions).
If such a constraint is denoted by $\epsilon$ and assuming the throughput equation \eqref{eq:5}, we have the following:
\begin{equation}
\begin{aligned}\label{eq:7}
& \underset{(\beta,m)}{\text{max}}
& & \frac{\log(1+\beta)}{1+\bar{m}}\times\left(1-P_\textrm{out}^{1+m}\right)  \\ 
& \text{subject to}
& & P_\textrm{out}^{1+m} \leq \epsilon 
\end{aligned},
\end{equation} 
where the SIR threshold $\beta>0$ and the number of allowed retransmissions $m \in \mathbb{N}$ are the design variables.

\begin{lemma}
The throughput $T$ in \eqref{eq:5} is a function of the variables $m>0$ and $\beta>0$, i.e.  $T = f (\beta, m)$. The function $f$ is then concave with respect to $\beta$ if $\frac{\partial^2 T}{\partial \beta^2} < 0$.
\end{lemma}

\begin{IEEEproof}
As $m$ and $\beta$ are strictly positive variables and function $T$ is twice differentiable in terms of $\beta$, then $T$ is concave if and only if $\frac{\partial^2 T}{\partial \beta^2} < 0$.
\end{IEEEproof}

\begin{lemma}\label{lemma_beta-cons}
If $T$ is in the region where $\frac{\partial^2 T}{\partial \beta^2} < 0$, then  $\beta^\ast$ represents the value of $\beta$ that maximizes the throughput and is computed as:
\begin{equation}\label{eq:8}
\beta^\ast=\left(- \frac{1}{k\lambda} \log{\left (1 - \epsilon^{\frac{1}{m + 1}} \right )}\right)^{\frac{\alpha}{2}}.
\end{equation}
\end{lemma}

\begin{IEEEproof}
We follow the derivation of \cite[Prop.1]{pedro2}.
First, we consider the equality in the optimization constraint: $P_\textrm{out}^{1+m} = \epsilon$
The average number of transmissions attempts $1+\bar{m}$ is:
\begin{equation}\label{eq:4}
1+\bar{m}= \sum_{n=0}^{m} \; P_\textrm{out}^n \approx \frac{1-P_\textrm{out}^{1+m}}{1-P_\textrm{out}} \approx \frac{1-\epsilon}{1-\epsilon^{\frac{1}{1+m}}}.
\end{equation}

By inserting \eqref{eq:6} and \eqref{eq:4} into \eqref{eq:5}, we have
\begin{equation}\label{eq:9}
T = \log(1+\beta) \left(1 - \epsilon^{\frac{1}{m + 1}}\right). 
\end{equation}

Eq. \eqref{eq:8} is then attained by solving the derivative equation $\partial T /\partial \beta = 0$, whose solution is $\beta^\ast$. It should be noted that \eqref{eq:4} that is used here is an approximation of \cite[§17]{nardellimult2009i} which proved to be a very good approximation for our studied $m$ and $\epsilon$ range.
\end{IEEEproof}

\begin{proposition}
The maximum allowed number of retransmissions $m^\ast$ that maximizes the link throughput is given by:
\begin{align}\label{eq:10}
 m^\ast=  \max\limits_{m \in \mathbb{N}}  & \;\; \log\left (- \frac{1}{k \lambda} \log\left (1 -\epsilon^{\frac{1}{m + 1}}  \right ) \right ) + \nonumber \\ 
 &\alpha  \left(- \frac{1}{k\lambda}\right)^{\frac{\alpha}{2}} \; \frac{ \left(\log\left (1 - \epsilon^{\frac{1}{m + 1}} \right ) \right )^{\frac{\alpha}{2} - 1}}{2 - \frac{2}{k\lambda}  \left (\log\left (1 - \epsilon^{\frac{1}{m + 1}} \right ) \right )^{\frac{\alpha}{2}}} .
\end{align}
\end{proposition}

\begin{IEEEproof}
From (\ref{eq:9})  and (\ref{eq:8}), we find $T$ as a function of $m$ considering  $\beta^*$.
The optimal throughput $T^*$ in terms of both $m$ and $\beta$ is then given by the value of $m$ that maximizes the throughput, which is given in \eqref{eq:10}.
\end{IEEEproof}

\begin{remark} The maximum number of retransmissions $m$ is a natural number that is usually small, which makes the evaluation of \eqref{eq:10} computationally simple.

\end{remark}

\subsection{Unconstrained optimization}

We now turn our attention to the unconstrained optimization of the link throughput.
Differently from the previous subsection, the optimization problem does not involve a maximum acceptable error rate and therefore no outage probability. Hence, in this case, retransmissions become useless as far as the SIR constraint $\beta$ is unbounded.   
In this case, the optimization problem is the following.
\begin{align}\label{eq:un}
\underset{\beta}{\text{max}}\quad
\log(1+\beta) e^{- k \lambda  \beta^{2/\alpha}}.
\end{align}

\begin{lemma}\label{lemma:uncon}
The throughput given in \eqref{eq:un} is concave with respect to $\beta$ if $\frac{\partial^2 T}{\partial \beta^2} < 0$.
In this case, the value of $\beta$ that maximizes the throughput is denoted by  $\beta^*_\textup{un}$ and is computed as the solution of the following equality:

\begin{equation}\label{eq:op}
\alpha\beta=2\beta^{\frac{2}{\alpha}} k (1+\beta) \log (1+\beta).
\end{equation}
	
\end{lemma}

\begin{IEEEproof}
If $\frac{\partial^2 T}{\partial \beta^2} < 0$, then \eqref{eq:un} can be solved by finding the optimal $\beta$ as the solution of the equality $\frac{\partial T}{\partial \beta}=0$, which is a transcendental equation so $\beta^*_\textup{un}$ requires numerical solution.
\end{IEEEproof}

\begin{proposition}
	\label{prop:opt}
	The optimal throughoput $T^*$ achieved by constrained and unconstrained optimization problems presented in this section coincides.
\end{proposition}

\begin{IEEEproof}
	Lemma \ref{lemma_beta-cons} assumes that $P_\textrm{out}^{1+m} = \epsilon$ to derive the relation between $m^*$ and $\beta^*$ from \eqref{eq:9}.
	Then, we also have
	\begin{align}\label{eq:x}
		T &= \log(1+\beta) \left(1 - \epsilon^{\frac{1}{m + 1}}\right) \nonumber\\
		&= \log(1+\beta) \left(1 - P_\textrm{out}\right)\nonumber\\
		&= \log(1+\beta) e^{- k \lambda  \beta^{2/\alpha}},
	\end{align}
	which is the unconstrained throughput formula given in \eqref{eq:un}.
	Therefore, the optimal throughput $T^*$ achieved by jointly setting  $m^*$ and $\beta^*$ is the same as the one achieved by $\beta^*_\textup{un}$.
\end{IEEEproof}

\begin{remark} Both solutions involve numerical computation. However, the constrained optimization needs a numerical search from the set of natural numbers, while the unconstrained one is obtained from a transcendental equation (whose solution is in the set of positive real numbers).
In this case, we may infer that the constrained optimization may provide a less complex way to solve the unconstrained optimization.

\end{remark}

\section{Numerical results}
\label{sec:res}

We present here the numerical results considering the following (arbitrary) setting: reference sensor-controller distance $r_0=1$ and path-loss exponent $\alpha=4$; the required error rate $\epsilon$ and the density of interferers $\lambda$ are the input parameters that their effects are analyzed.

\begin{figure}[!t]
	\vspace*{-5mm}
	\includegraphics[width=\columnwidth]{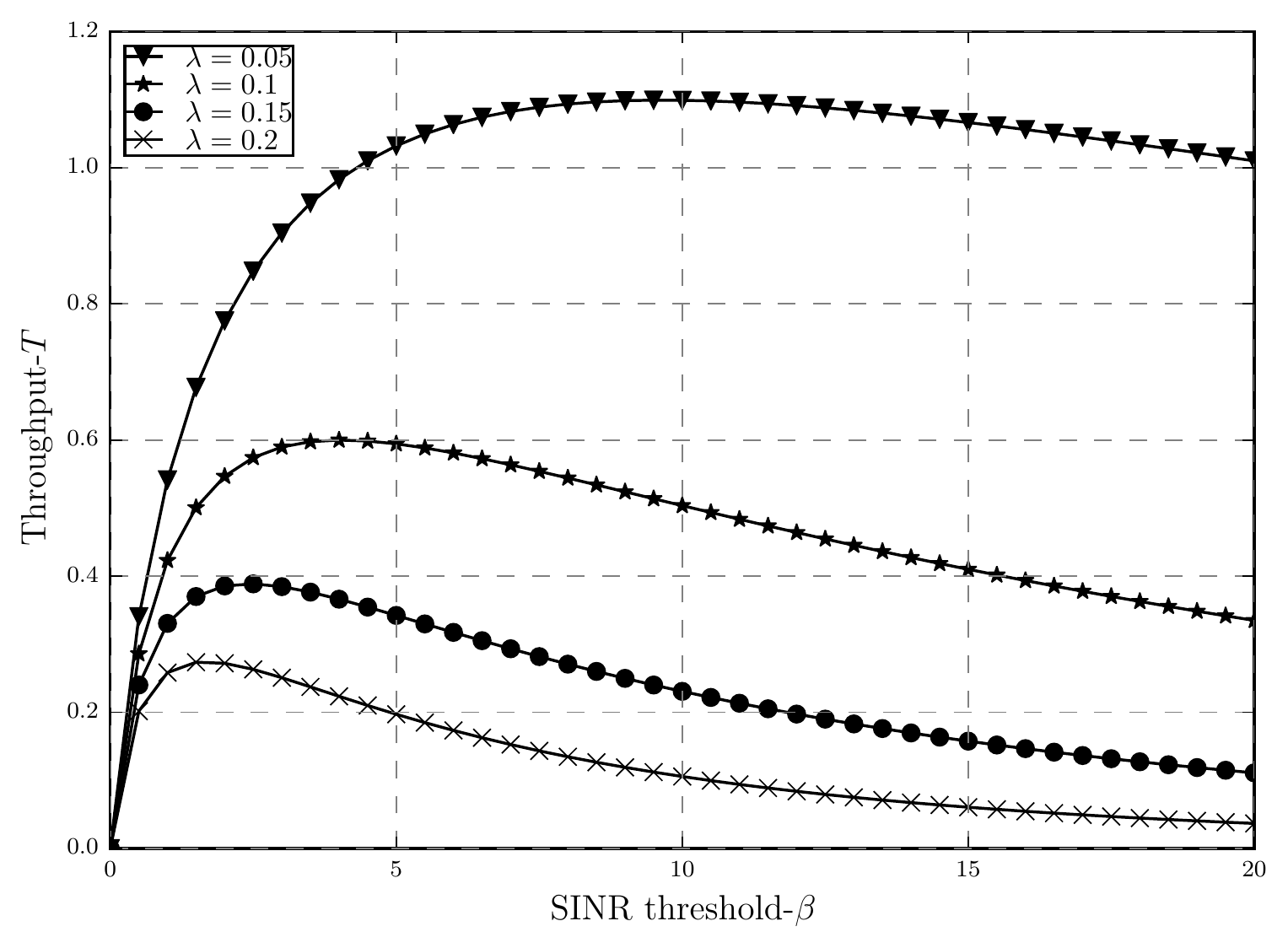}
	\caption{Throughput $T$ versus the SINR threshold $\beta$ for $\alpha=4$, $r_0=1$ and different densities $\lambda$.}
	\label{fig:0}\vspace{-3mm}
\end{figure}
\begin{figure}[!t]
	\vspace*{-5mm}
	\includegraphics[width=\columnwidth]{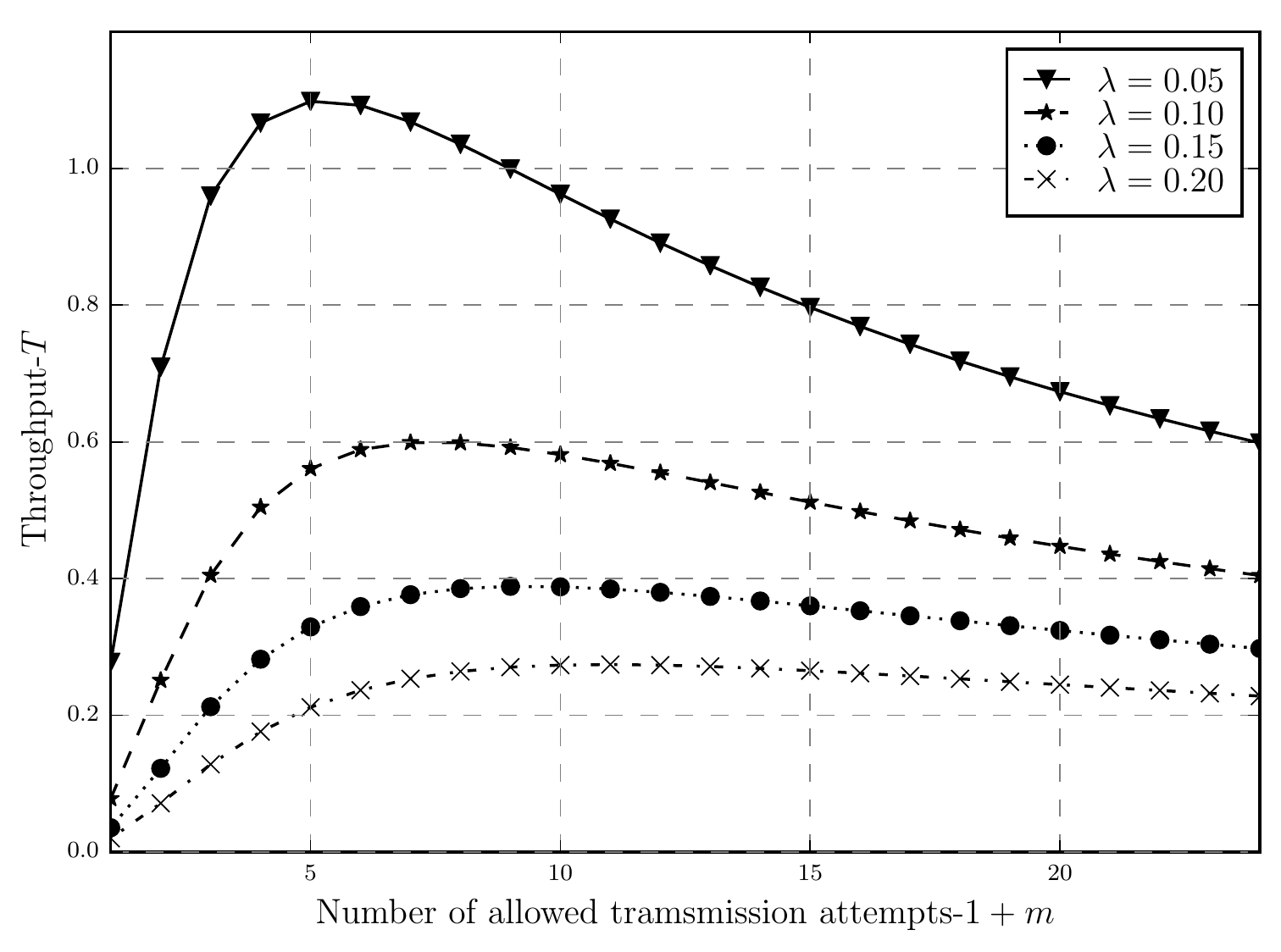}
	\caption{Throughput $T$ versus the maximum number of allowed transmissions attempts $1+m$ for $\alpha=4$, $r_0=1$ and different densities $\lambda$.}
	\label{fig:2}
\end{figure}

Fig. \ref{fig:0} and \ref{fig:2} shows the behavior of the link throughput $T$ as a function of the SINR threshold $\beta$ and the maximum number of allowed retransmissions $m$ for different $\lambda$ and $\epsilon = 0.02$ (98\% of success after retransmissions) respectively. We can see that as $\beta$ increases in Fig. \ref{fig:0}, the link throughput gets higher for different values of $\lambda$ until at some point it reaches the trade of point after which, it starts decreasing. This is due to the fact that although by increasing $\beta$ the system will have a more efficient transmission, it will also increases the outage events which will lead to a decrease in the link throughput. The same twofold effect can also be explained for Fig. \ref{fig:2}.
While increasing $m$ implies greater values of $\beta$
which  allows for higher spectral efficiencies in each transmission attempt -- greater $\log(1+\beta)$, it also decreases the chance of the message being successfully decoded in a single attempt, increasing $P_\textup{out}$.
These trade-offs are captured by the proposed constrained optimization, whose optimal point $T^*$ is found by setting $m^*$ and $\beta^*$.

For example, when $\lambda=0.05$, the best design setting is $m^*=4$, or $1+m^*=5$, and $\beta^*=6.14$.
This leads to $300\%$ improvement compared with the optimal throughput achieved without retransmissions (i.e. $m=0$).
Likewise, setting $m^*=4$ achieves about $200\%$ more throughput when a large number $m$ is allowed.

This effect is similar when different densities $\lambda$ are assumed.
Note that $\lambda$ reflects the number of active transmitters in the licensed network.
Consequently, the greater the $\lambda$, the higher the interference experienced at the unlicensed link.
This then decreases the optimal throughput $T^*$, which  is nevertheless obtained with a reasonable low value; for example, $m^*= 10$ when $\lambda = 0.2$, which evinces that the numerical search for such optimal value is computationally cheap.

\begin{figure}[!t]
	\vspace*{-5mm}
	\includegraphics[width=\columnwidth]{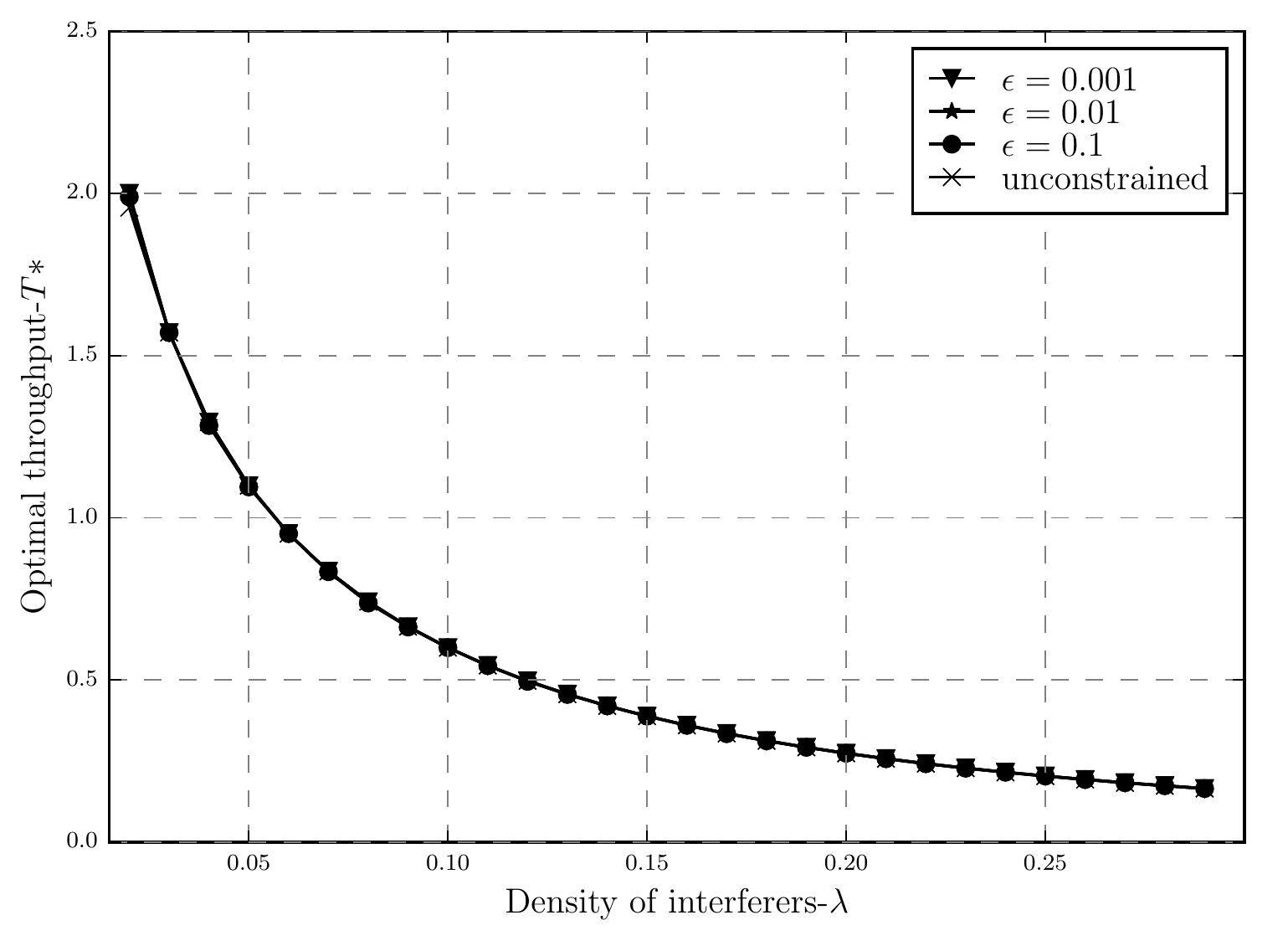}
	\caption{Optimal Throughput $T^\ast$ versus the density of interferers $\lambda$ for both unconstrained and constrained optimizations where $\alpha=4$, $r_0=1$ with unlimited number of retransmissions.}
	\label{fig:5}\vspace{-3mm}
\end{figure}

\begin{figure}[!t]
	
	\includegraphics[width=\columnwidth]{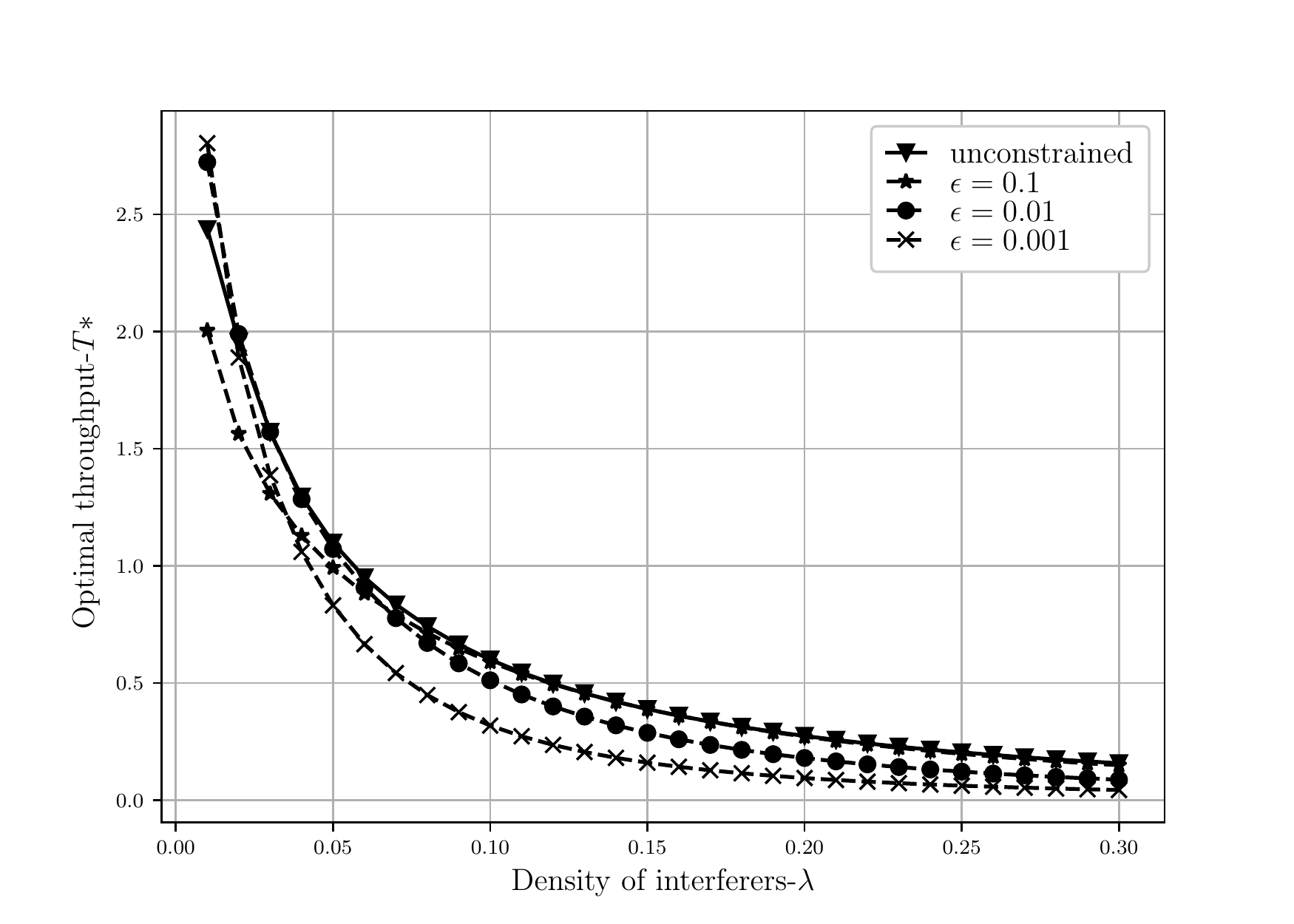}
	\caption{Optimal Throughput $T^\ast$ versus the density of interferers $\lambda$ for both unconstrained and constrained optimizations where $\alpha=4$, $r_0=1$ with limited number of retransmissions ($m=5$).}
	\label{fig:5_2}\vspace{-3mm}
\end{figure}

\begin{figure}[!t]
	\vspace*{-5mm}
	\includegraphics[width=\columnwidth]{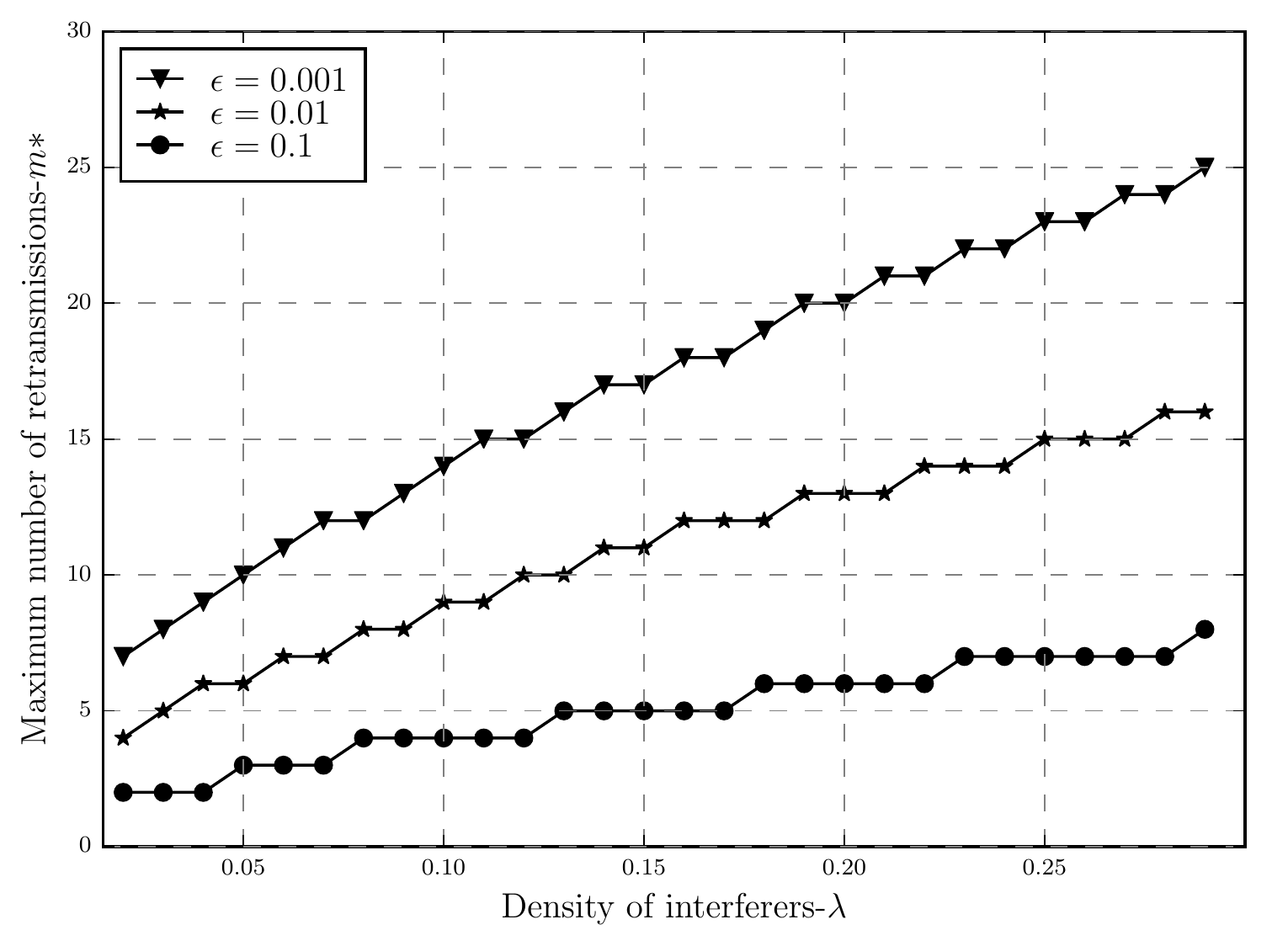}
	\caption{Maximum number of retransmissions $m$ versus density $\lambda$ for $\alpha=4$, $r_0=1$ and different error rate requirements $\epsilon$.}
	\label{fig:4}\vspace{-3mm}
\end{figure}

Fig. \ref{fig:5} and \ref{fig:5_2} show the optimal constrained and unconstrained throughputs $T^*$ as a function of $\lambda$; with unlimited and limited number of retransmissions respectively. In this case, we consider different error rate requirements $\epsilon$. 
As predicted by Proposition \ref{prop:opt}, the optimal throughput $T^*$ is independent of $\epsilon$ and can always achieve the unconstrained optimal value while the number of retransmissions is not limited. On the other hand, if the maximum number of retransmissions are limited as in Fig. \ref{fig:5_2}, the error rate constraints will affect the optimal throughput and it can not achieve the unconstrained $T^*$ anymore. In Fig. \ref{fig:5}, system is allowed unlimited retransmissions hence, as the  $\epsilon$ requirement gets stricter, the system uses more retransmissions to compensate for that and reach the unconstrained $T^*$, while in \ref{fig:5_2} since $m^*$ is limited and $m^*=5$, this compensation will not happen anymore, thus, resulting in a lower $T^*$ for the same $\lambda$. It should also be noted that the lower the $\epsilon$, the better the approximated \eqref{eq:4} works. That is the reason behind the throughput curve behavior for $\epsilon=0.1$ for a small portion of the $\lambda$ range. 

This result indicates the importance of using retransmissions to regulate the spectral efficiency (through the SIR threshold) in order to achieve the highest possible link throughput for a given density of interferers $\lambda$.
To better understand how the retransmissions work to optimize the link throughput, we show in Fig. \ref{fig:4} how the number of maximum retransmissions $m^*$ changes with $\lambda$ for different $\epsilon$.
As expected, the number of retransmissions that leads to the optimal throughput increase as either the density of interferers increases or the quality requirement $\epsilon$ gets stricter. Comparing this figure with Fig. \ref{fig:5} and \ref{fig:5_2} and considering $\lambda=0.1$ and $\epsilon=0.01$ as an example, we can see that the optimal throughput that the system can reach while $m^*$ is limited to $5$ is $T^*=0.45$ and 4 more retransmissions are needed in order to reach the unconstrained $T^*$ as it happens in Fig. \ref{fig:5}. 
Once the value $m^*$ is set, the SIR threshold $\beta^*$ -- which leads to the spectral efficiency $\log(1+ \beta^*)$ -- comes straight from \eqref{eq:8}. Fig. \ref{fig:4} can also be considered a representative of the delay of the network if it is expressed in terms of the number of the packets for different network densities. Considering $\lambda=0.1$ for instance, while the error rate constraint of the network is $\epsilon=0.01$, the delay would be equal to $1+m^*=10$ times the packets' transmission time.

In this case, the pair $(\beta^*, m^*)$ that leads to $T^*$ constrained by $\epsilon$ is explicitly obtained.
Such an optimal throughput (interestingly) coincides with the one obtained by the unconstrained optimization via the numerical solution of \eqref{eq:op}, which defines $\beta^*_\textup{un}$ and spectral efficiency $\log(1+ \beta^*_\textup{un})$.
Although throughput-maximizing, $\beta^*_\textup{un}$ usually leads to relatively high outage probabilities; for example, $P_\textup{out}=0.5$ for $\lambda=0.05$.
The proof of Proposition \ref{prop:opt} tells that retransmissions act to allow for higher outage probabilities at each transmission, respecting after all attempts the (generally strict) constraint $\epsilon$.

\section{Conclusion}
\label{sec:concl}

This paper assessed an unlicensed spectrum access scenario for a class of wireless systems where unlicensed users employ the uplink channel of the licensed network in a way that the interference caused by them can be neglected. 
Our results showed that even with very stringent reliability error rate requirements, it is possible to achieve the optimal unconstrained link throughput.
This can only be achieved by allowing for a limited number of retransmissions of messages decoded in outage.
The proposed analysis derived the combination of SIR threshold and allowed retransmissions that reaches the optimal link throughput when the interfering nodes from the licensed network are modeled as a Poisson point process.
Although retransmission increases the system throughput, it will also affect the delay of the network, hence, it is important to find a trade off between these different and competing network requirements. We expect to extend these results by including such trade off analysis in addition to matters related to secrecy \cite{pedro+sec} and the specific signal under consideration (e.g. \cite{tome}) as well.

\vspace{-3mm}
\section*{Acknowledgments}

This work is partially supported by Aka Project SAFE (Grant n.303532) and Strategic Research Council/Aka BCDC Energy (Grant n.$292854$).

\bibliographystyle{IEEEtran}

\begin{thebibliography}{10}
\providecommand{\url}[1]{#1}
\csname url@samestyle\endcsname
\providecommand{\newblock}{\relax}
\providecommand{\bibinfo}[2]{#2}
\providecommand{\BIBentrySTDinterwordspacing}{\spaceskip=0pt\relax}
\providecommand{\BIBentryALTinterwordstretchfactor}{4}
\providecommand{\BIBentryALTinterwordspacing}{\spaceskip=\fontdimen2\font plus
\BIBentryALTinterwordstretchfactor\fontdimen3\font minus
  \fontdimen4\font\relax}
\providecommand{\BIBforeignlanguage}[2]{{%
\expandafter\ifx\csname l@#1\endcsname\relax
\typeout{** WARNING: IEEEtran.bst: No hyphenation pattern has been}%
\typeout{** loaded for the language `#1'. Using the pattern for}%
\typeout{** the default language instead.}%
\else
\language=\csname l@#1\endcsname
\fi
#2}}
\providecommand{\BIBdecl}{\relax}
\BIBdecl

\bibitem{iotserv}
A. Al-Fuqaha et al. "Internet of Things: A Survey on Enabling Technologies, Protocols, and Applications," in \textit {IEEE Communications Surveys and Tutorials}, vol. 17, no. 4, pp. 2347-2376, Fourthquarter 2015.




\bibitem{ak}
I. F. Akyildiz et al. "NeXt generation/dynamic spectrum access/cognitive radio wireless networks: A survey." \textit{ Computer networks}, vol. 50, no. 13, pp. 2127-2159, 2006.

\bibitem{pdnd}
P. H. J. Nardelli et al. "Throughput analysis of cognitive wireless networks with Poisson distributed nodes based on location information." \textit {Ad Hoc Networks}, vol. 33, pp. 1-15, 2015.


\bibitem{pedro2}
P. H. J. Nardelli et al. "Maximizing the link throughput between smart meters and aggregators as secondary users under power and outage constraints." \textit {Ad Hoc Networks}, vol. 41, pp. 57-68, 2016.

\bibitem{tome}
M. C. Tomé et al. "Joint sampling-communication strategies for smart-meters to aggregator link as secondary users."  \textit{2016 IEEE International Energy Conference (ENERGYCON)}, Leuven, 2016, pp. 1-6.

\bibitem{mar}

P. H. J. Nardelli et al. "Optimal Transmission Capacity of Ad Hoc Networks with Packet Retransmissions," in \textit{ IEEE Transactions on Wireless Communications}, vol. 11, no. 8, pp. 2760-2766, August 2012.

\bibitem{pedro+marios}

P. H. J. Nardelli et al. "Throughput Optimization in Wireless Networks Under Stability and Packet Loss Constraints," in \textit{IEEE Transactions on Mobile Computing}, vol. 13, no. 8, pp. 1883-1895, Aug. 2014.

\bibitem{joint}
J. Wildman et al. "On the Joint Impact of Beamwidth and Orientation Error on Throughput in Directional Wireless Poisson Networks," in \textit{IEEE Transactions on Wireless Communications}, vol. 13, no. 12, pp. 7072-7085, Dec. 2014.


\bibitem{haen}
M. Haenggi. Stochastic geometry for wireless networks. \textit{Cambridge University Press}, 2012.

\bibitem{weber}
S. Weber et al. "An Overview of the Transmission Capacity of Wireless Networks," in \textit{IEEE Transactions on Communications}, vol. 58, no. 12, pp. 3593-3604, December 2010.

\bibitem{ptp}
F. Baccelli et al. "Interference networks with point-to-point codes," 2011 \textit{IEEE International Symposium on Information Theory Proceedings}, pp. 435-439, St. Petersburg, 2011.

\bibitem{pedro+sec}
P. H. J. Nardelli et al. "Throughput maximization in multi-hop wireless networks under a secrecy constraint." \textit{Computer Networks}, vol. 109, pp. 13-20, 2016.


\bibitem{nardellimult2009i}
P. H. J. Nardelli et al. Multi-hop aggregate information efficiency in wireless ad hoc networks. In: Communications, 2009. ICC'09. IEEE International Conference on. IEEE, 2009. p. 1-6.





\end{thebibliography}

\end{document}